# Structure and magnetocaloric properties of $La_{1-x}K_xMnO_3$ manganites

A.M. Aliev[*,1], A.G. Gamzatov[**,1], A.B. Batdalov[1], A.S. Mankevich[2], I.E. Korsakov[2]

[1]*Institute of Physics of Daghestan Scientific Center of RAS, Makhachkala, 367003, Russia*
[2]*Moscow State University, 119899, Moscow, Russia*

[*]E-mail: lowtemp@mail.ru
[**]E-mail: gamzatov_adler@mail.ru

A technology of obtaining the single-phase ceramic samples of $La_{1-x}K_xMnO_3$ manganites and the dependence of their structural parameters on the content of potassium has been described. Magnetocaloric effect (MCE) in the obtained samples has been measured by two independent methods: by classical direct methodic and by a method of magnetic field modulation. The values of MCE obtained by both methods have been substantially differed. The explanation of the observed divergences is given. The correlation between the level of doping and MCE value has been defined. The value of $T_C$ determined by the MCE maximum has been conformed to the literature data received by other methods.

**Introduction**

Magnetic cooling based on the magnetocaloric effect is simple, convenient, and promising technology of cooling. Therefore, the materials with large magnetocaloric effect are actively searched and examined [1]. The great values of MCE are observed in the rare-earth elements and the alloys based on them, owing to the large magnetic moments of the atoms in these elements [1-4]. Of late years the new magnetocaloric materials (manganites, Heusler's alloys) arouse interest. The maxima of MCE are great in these compounds and revealed near room temperatures. A substantial advantage of these materials is an ability to adjust the temperature of phase transitions over wide limits, including a region of room temperatures by means of a change in the concentration of the chemical elements in them.

The doped manganites have a number of interesting physical properties; the colossal magnetoresistance, phase separation, co-existence of different types of orderings, etc. Most significant values of MCE are observed in the majority of manganite compounds. Physical and magnetocaloric properties are extensively studied for manganites doped with divalent alkaline earth-elements [3, 5-9]. The manganites doped with univalent elements have been begun to study relatively recently; in this case, the researchers pay the greatest attention to the system of $La_{1-x}Ag_xMnO_3$, which is characterized by rather substantial values of MCE and temperatures of phase transitions convenient for the practical applications [10-12].

The system of $La_{1-x}K_xMnO_3$ has not been studied minutely; in any event the works devoted to the physical properties of this system are very few [13-19], and publications providing the data about magnetocaloric properties are even less [20-23]. Moreover, MCE in the majority of these researches is evaluated intermediary, by measuring the magnetization, using the Maxwell correlation $\Delta S = \int_0^H (\partial M/\partial T)_H dH$ ($\Delta S$ is a change of magnetic entropy, $M$ is magnetization).

This work is devoted to the research of the magnetocaloric properties of $La_{1-x}K_xMnO_3$ manganites ($x = 0.05, 0.1, 0.11, 0.13, 0.15, 0.175$) in magnetic fields from 500 Oe to 11 kOe.

**Samples and experiment**

The $La_{1-x}K_xMnO_3$ samples were synthesized using a Pechini route from $Mn(NO_3)_2$, $La(NO_3)_3$, and $K_2CO_3$ ("Reakhim", "pure for analysis" grade). A stoichiometric amount of $K_2CO_3$ together with citric acid and ethylene glycol were added to a solution of Mn and La nitrates. The solution was evaporated at 60 °C until gelatinous state, and the gel was decomposed at 600 °C for 5 hours in air. The obtained powder was pressed into pellets which were placed into powder of the same composition to prevent potassium evaporation and annealed at 1000 °C for 30 hours in air.

All solid solutions $La_{1-x}K_xMnO_3$ (LKMO) have rhombohedraly distorted perovskite-type structure. The powder XRD patterns of LKMO show that degree of such distortion decreased with increase of potassium concentration ($x$), what is apparent through decrease of splitting in the pseudocubic reflections (Fig. 1) and reduction of rhombohedral angle $α$ (Fig. 2) with increasing of potassium concentration. An approach of the rhombohedral angle $α$ to 60° indicates a decreasing of rhombohedral distortion. Sample with $x=0.2$ consists of a perovskite-like solid solution (La,K)$MnO_3$ and a K-rich form of birnessite $K_yMnO_2$. In our previous work was shown that samples with nominal composition $La_{1-x}K_xMnO_3$ ($x>0.18$) consisted of three main phases: the $La_{0.82}K_{0.18}MnO_3$ perovskite, K-birnessite $K_{0.49}MnO_2$ and minor amount of potassium manganite $K_2MnO_4$ [24]. The structure of LKMO can be represented in both hexagonal and rhombohedral unit cells. The cell parameters of LKMO are determined from XRD patterns get from powders with addition of high purity silicon as internal standard. Results for unit cell parameters are presented in Table 1 and Figure 2.

All powder XRD patterns of LKMO solid solutions were collected at room temperature. In this case, because of different temperatures of the ferromagnetic ordering, one part of the samples ($x<0.12$) measured in the paramagnetic state and the other part ($x≥0.12$) in the ferromagnetic state. It is possible that lattice of solid solutions with $x≥0.12$ was distorted due to spontaneous magnetostriction, that causes the feature of potassium content depending of the lattice parameters of LKMO solid solutions.

A cation composition of the prepared samples was verified by an atomic-emission spectrometry of induced coupled plasma and mass-spectrometry of induced coupled plasma. Table 2 summarizes the results of the chemical analysis of samples sintered at 1000 °C. As can be seen, the defined composition for all samples is practically equal to the nominal one.

As noted above, the magnetocaloric effect was measured by two methods. The first is the classical direct method, in which the adiabatic temperature change of the sample is recorded with a change in magnetic field. The essential defect of this method is labour- intensiveness, and also the need of massive samples, which are required for exact measuring of temperature change. The second method, proposed recently in [26], is a modulation method, which has a number of substantial advantages. First of all, it is a great sensitivity of the method, what allows of measuring the MCE with slight changes in magnetic field. The massive samples are not required for the study. The characteristic sizes of the investigated samples have been 3x3x0.3 mm3, though the method allows to measure super-small samples of 10-20 μm in thickness. The measurements have been carried out in quasi-adiabatic regime, which considerably reduces the time of the experiment. The average time required to measure the MCE in the temperature range of 100 K has been about 2 hours, which is in several times less than required by the standard method.

**Results and discussion**

Figure 3 shows the temperature dependence of the MCE in samples of $La_{1-x}K_xMnO_3$, measured by a modulation method with the amplitude changes in magnetic field of 500 Oe. As seen in Figure 3, the temperatures of effect maxima grow with increasing in concentration of potassium from $T$= 186 K for $x$=0.05 to $T$=333 K for $x$=0.175. If we exclude the composition of $x$=0.05, where the MCE value is substantially less and observed at low temperatures, then the temperature interval, where the effect maxima fall on, is equal to 270-330 K for $x \geq 0.1$, i.e. it covers the room temperature. As a factor defining the MCE value in manganites is the quantitative correlation between the different valence manganese ions of $Mn^{3+}$ and $Mn^{4+}$, which realize the double exchange interaction: the more such pairs, the larger the magnetization and magnetic entropy change, i.e. the MCE. Meanwhile, the experimental facts show that the greatest values of MCE in manganites have been observed in the $Mn^{3+}/Mn^{4+} \approx 7:3$ [27-30]. This correlation between the manganese ions should be at $x$=0.3, assuming that each introduced divalent metal ion produces one ion of $Mn^{4+}$. Such correlation for univalently substituted manganites will occur when $x$=0.15 [31, 32], as in this case for every ion of univalent metal two ions of $Mn^{3+}$ must pass into the state of $Mn^{4+}$ in order to save the charge balance. The analysis of our experimental data shows that for the system $La_{1-x}K_xMnO_3$ the MCE value grows with increasing of $x$, reaches maximum at $x$=0.13, and further increasing in $x$ leads to decrease of effect magnitude. It can mean that with $x$>0.13 a part of the newly formed ions of $Mn^{4+}$ participates in super-exchange interaction of $Mn^{4+}$–O–$Mn^{4+}$, what is favorable to antiferromagnetic ordering and therefore the volume fraction of ferromagnetic phase has been decreased in the model.

A measuring of MCE is a sufficiently precise indirect method to determine the $T_C$ of the magnetically ordered materials. According to [33], the MCE maximum is observed in a point, at which the heat capacity in the magnetic field is equal to the heat capacity in the zero-field, i.e., $C_P(T, H)=C_P(T, 0)$ and this point always locates near or above the heat capacity maximum temperature. If we plot the dependence of $T_{max}$ by the different values of magnetic field $H$, then the $T_{max}$ value of magneto-caloric effect extrapolated to the zero-field coincides with $T_C$ of this material. Actually, it means that $T_{max}$ of magneto-caloric effect in the low magnetic fields is equal to $T_C$. The $T_C$ values, obtained thereby, for fields $\Delta H$=500 Oe are shown in Table 3, and the experimental data for one sample are presented in Figure 4. For comparison here are shown the $T_C$ values, obtained on the basis of the data analysis $C_P(T)$. As can be seen from the Tables, the $T_C(x)$ values determined by both methods approximately coincide with each other and literature data [13, 21, 23]. They have a general tendency: $T_C$ grows with increase in level of alloying for entire interval of $x$=0.05-0.175. At the same time the certain data testify to that the dependence of T(x) passes through the maximum in some univalent substituted manganites [32, 34, 35]. The observed differences can be caused by the large difference in ionic radii of La$^{3+}$ и K$^+$. The doping level of $x$ effects on $T_C$ by changing the average ionic radius of A-cations $\langle r_A \rangle$ and occurring a disorder, due to the difference in ionic radii of cations in the A-sublattice, defined as $\sigma^2 = \sum x_i^2 r_i^2 - \langle r_A \rangle^2$ ($x_i$ is a concentration of $i$- cation, $r_i$ is the radius of $i$- cation). Both of these factors effect on $T_C$ differently: the growth of disorder parameter $\sigma^2$ leads to reduction of the exchange interaction and a decrease in $T_C$, while a growth of $\langle r_A \rangle$ increases the $T_C$ [36, 37].

The experimentally observed growth in $T_C$ with increasing $x$ can only mean that the degree of effect of $\langle r_A \rangle$ on the $T_C$ in the interval of potassium concentrations exceeds the influence connected with local distortions of crystal lattice determined by $\sigma^2$. Actually, a great difference in ionic radii of La and K ($r_{La^{3+}} = 1.18 \dot{A}, r_{K^+} = 1.52 \dot{A}$) leads to a rapid growth of $\langle r_A \rangle = (1-x) r_{La^{3+}} + x \cdot r_{K^+}$ during alloying, ractification of Mn-O-Mn bond, amplification of the exchange interaction, and therefore increase in $T_C$.

The maximum of MCE values vary within the range from $\Delta T$=0.027 K in La$_{0.95}$K$_{0.05}$MnO$_3$ to $\Delta T$=0.136 K in La$_{0.87}$K$_{0.13}$MnO$_3$ with a change in magnetic field of 500 Oe. The linear extrapolation of MCE values, obtained at $\Delta H$ = 500 Oe, to strong magnetic fields ($\Delta H$=10 kOe) provides $\Delta T$=2.72 K for La$_{0.87}$K$_{0.13}$MnO$_3$.

This value approximately corresponds to the maximum magnitude of magneto-caloric effect in gadolinium with the same fields [1]. MCE measurements by the direct classical method at $\Delta H$=10 kOe have shown substantially lower values of MCE for the same samples (Fig. 5). It's not reason to attribute this mistake for the modulation method, as extrapolated data for gadolinium, obtained by modulation method in low fields, approximately coincide with MCE values in strong magnetic fields.

Therefore, it is possible to assume that the difference in the values of MCE, obtained by linear extrapolation from one side and measured in the strong magnetic fields from another side, are connected with the nonlinear field dependence of MCE of the investigated models in field ranges of 0-10 kOe.

Indeed, a nonlinear connection between $\Delta S_{max}$ and $\Delta H$ in a wide range of magnetic fields up to 60 kOe has been revealed in $La_{0.7}Ca_{0.3-x}K_xMnO_3$ [17] and $La_{0.7}Ca_{0.2}Sr_{0.1}MnO_3$ [27]. It may mean that the extrapolation procedure used for manganites $La_{1-x}K_xMnO_3$ is not entirely correct, and it is necessary to specify the nature of the $\Delta T = f(\Delta H)$ dependence for the objective evaluations.

For this purpose magneto-field dependences of MCE have been investigated for two samples - $La_{0.87}K_{0.13}MnO_3$ and $La_{0.85}K_{0.15}MnO_3$ close to $T_C$ (Fig. 6). Obtained data are approximated by expression $\Delta T = \alpha \cdot H^{0.91}$ ($\alpha$ is constant of proportionality) for both of models. At the same time the $\Delta S_M = f(H)$ dependence calculated in the approximation of the mean-field theory near $T_C$ has the following form [38]: $\Delta S_M = \beta \cdot H^{2/3}$ ($\beta$ is a parameter independent from temperature and field). In recent researches in [39] it was shown that $\Delta S_M \sim H^n$, where $n=1$ at $T \ll T_C$, $n=2$ at $T \gg T_C$ and $n=0.75$ near $T_C$.

For comparison of experimentally determined adiabatic temperature change in $\Delta T(H)$ with the calculated $\Delta S_M(H)$ values it must be used the relation binding $\Delta T$ and $\Delta S_M$ [33]: $\Delta T = -(T \Delta S_M / C_H)$. This expression is valid away from $T_C$, where the $C_P$ does not depend on H and near $T_C$ in low magnetic fields, what agrees with our results. It follows that field dependence $\Delta S_M$ and $\Delta T$ must have identical character, and if one follows the results in [39], then $\Delta T \sim H^{0.75}$, while experiment has shown the $\Delta T \sim H^{0.91}$. The observed differences in exponent can be easily explained if we consider the dependence of the heat capacity on the magnetic field: $C_P$ decreases near $T_C$ with increase of the field [20], which must lead to increase in dependence of $\Delta T$ on $H$. Using the expression $\Delta T = \alpha \cdot H^{0.91}$, resulting from measurements in the low fields for the MCE extrapolation to the strong fields, we derive $\Delta T = 2.05$ K and 1.66 K for $La_{0.87}K_{0.13}MnO_3$ and $La_{0.85}K_{0.15}MnO_3$, respectively, at $\Delta H = 10$ kOe. These values already more less than the values, obtained by linear extrapolation, and close to the values, presented in the work [22]. Hence, it follows that the real values of MCE in the strong magnetic fields are possible to receive by investigating field dependences of MCE in the low fields.

Let's compare the MCE values, obtained by different methods. We fail to compare the results of our researches in low fields with literature data, since there are no any data in literature of it. The data about direct measurements of adiabatic temperature change in the strong fields are also absent. There is only estimation of $\Delta T$ from the data on magnetization [21 - 23]. According to [22, 23], the MCE value with change of the field by 10 kOe are 1.79 K and 2.06 K for the compositions with $x=0.1$ and 0.15, respectively. In comparison with the results of our direct measurements by the classical method [20], these data are about twice as much. The distinction in MCE values, obtained by different methods can be explained as by various errors of used methods, so by the distinction in quality of the models through the different techniques of their preparing. Direct measurements always give the understated values relative to the basic values of MCE. It is caused by the fact that the basic parameter, determinating the accuracy of direct measurements, the heat insulation, cannot be reduced to zero, and the measured value of a temperature change will be less than the actual due to the heat dissipation. When the MCE is measured by the classical magnetic method, the rise time of the magnetic field usually takes several seconds, and the energy gained by the lattice from the magnetic field is partially evaporating for this time (because of bad vacuum, the leakage of heat over measuring wires, etc). In case of using the modulation method the rise (fall) time of the field is $t=(1/4\nu)$, where $\nu$ is the frequency of the voltage supplied to the electromagnet. For $\nu=0.3$ Hz, the rise time of field from zero to amplitude value will be $t\approx0.8$ s. When $\Delta H$ is low, $\Delta T$ will be low too (by 0.1 K and less at the amplitude of magnetic field 500 Oe). The low gradient of temperature between the sample and the environment means that a speed of heat dissipation will be slow. Moreover, for the registration the temperature oscillation is used a differential thermocouple, junction of which is flattened to the thickness of 3-5 μm, what strongly reduces a thermocouple persistence, and a large contact area substantially improves thermal contact. Therefore, a use of the modulation method will provide more precise, true MCE values.

The equally important advantageous aspect of the modulation method is that it allows not only to measure the MCE with high accuracy, but also to determine a type of magnetic ordering (ferromagnet, antiferromagnet), and to observe the evolution of a change in volume fractions, what was successfully demonstrated recently for $Pr_{1-x}Ag_xMnO_3$ [40]. A disadvantage, limiting a usage of this method, is a lack of the source of high-frequency magnetic fields. This disadvantage can be removed soon.

**Conclusion**

Thus, MCE in the manganites of $La_{1-x}K_xMnO_3$ is investigated by two independent methods: 1. classical straight method; 2. magnetic field modulation method. The advantages and disadvantages of each method and reasons for the observed differences in results are discussed. The $T_C$ values estimated on the basis of MCE data coincide with the data defined by other method. The MCE in strong magnetic fields is shown to be estimated by means of precise measurements of MCE in low fields.

**Acknowledgements**

This study was supported by the Russian Foundation for Basic Research (project no. 09-08-96533), the Program of the Physical Sciences Division of the Russian Academy of Sciences «Highly correlated electrons in solids and structures».

**Reference**

1. A. M. Tishin and Y. I. Spichkin, *The Magnetocaloric Effect and its Applications*, 1st ed. (Institute of Physics, New York, 2003).
2. V.K. Pecharsky, K.A. Gschneidner, Phys. Rev. Lett. **78**, 4494 (1997).
3. Manh-Huong Phan, Seong-Cho Yu, Journal of Magnetism and Magnetic Materials **308**, 325 (2007).
4. Antoni Planes, Lluís Mañosa and Mehmet Acet, Journal of Physics: Condensed Matter **21**, 233201 (2009).
5. M. Bejar, E. Dhahri, E.K. Hlil, S. Heniti, Journal of Alloys and Compounds **440**, 36-42 (2007).
6. P. Sarkar, P. Mandal, and P. Choudhury, Appl. Phys. Letters **92**, 182506 (2008).
7. A. Biswas, T. Samanta, S. Banerjee and I. Das, Journal of Physics: Condensed Matter **21**, 506005 (2009).
8. M.S. Reis, V.S. Amaral, J.P. Araujo, P.B. Tavares, A.M. Gomes, I.S. Oliveira, Phys. Rev. B. **71**, 144413 (2005).
9. W. Zhong, W. Chen, C. T. Au, Y. W. Du, Journal of Magnetism and Magnetic Materials **261**, 238 (2003).
10. T. Tang, K. M. Gu, Q. Q. Cao, D. H. Wang, S. Y. Zhang, Journal of Magnetism and Magnetic Materials **222**, 110 (2000).
11. K.Q. Wang, Y.X. Wang, A. Junod, K.Q. Ruan, G.G. Qian, M. Yu, L.Z. Cao. Phys. Status Solidi B **223**, 673 (2001).
12. I.K. Kamilov, A.G. Gamzatov, A.M. Aliev, A.B. Batdalov, A.A. Aliverdiev, Sh.B. Abdulvagidov, O.V. Melnikov, O.Y. Gorbenko, A.R. Kaul, Journal of Physics D: Appl. Phys. 40, 4413 (2007).
13. Soma Das, T.K. Dey, Physica B **381**, 280 (2006).
14. K.Q. Wang, Y.X. Wang, A. Junod, K.Q. Ruan, et al. Phys. Stat. Sol. (b) **223**, 673 (2001).
15. C. Shivakumara and Manjunath B. Bellakki, Bull. Mater. Sci. **32**, 443 (2009).
16. M. Bejar, R. Dhahri, E. Dhahri, M. Balli, E.K. Hlil, Journal of Alloys and Compounds, **442**, 136 (2007).
17. Soma Das and T.K. Dey, Bull. Mater. Sci. **29**, 633 (2006).
18. Wu Jian, Zhang Shi-Yuan. Chin, Phys. Lett. **21**, 382 (2004).
19. W. Zhong, W. Chen, W.P. Ding, et al. Eur. Phys. J. B **3**, 169 (1998).
20. I.K. Kamilov, A.G. Gamzatov, A.B. Batdalov, A.S. Mankevich, I.E. Korsakov, Physics of the Solid State, **52**, 789 (2010).

21. W. Zhong, W. Chen, W. P. Ding, N. Zhang, A. Hu, Y. W. Du, Q. J. Yan, Journal of Magnetism and Magnetic Materials **195**, 112 (1999).
22. Soma Das, T.K. Dey, Journal of Alloys and Compounds. **440**, 30 (2007).
23. Soma Das and T K Dey, J. Phys.: Condens. Matter 18, 7629 (2006).
24. J. Hadermann, A.M. Abakumov, S.V. Rompaey, A.S. Mankevich and I.E. Korsakov, Chemistry of Materials **21**, 2000 (2009).
25. S. Zouari, A. Cheikh-Rouhou, P. Strobel, M. Pernet, J. Pierre, Journal of Alloys and Compounds, **333**, 21 (2002)
26. A.M. Aliev, A.B. Batdalov, V.S. Kalitka. JETP Letters **90**, 663 (2009).
27. G.C. Lin, X.L. Yu, Q. Wei, J.X. Zhang, Materials Letters, 59, 2149-2152 (2005)
28. L.E. Hueso, P. Sande, D.R. Miguens, J. Rivas, F. Rivadulla, M.A. Lopez-Quintela, Journal of Applied Physics **91**, 9943 (2002).
29. W. Zhong, W. Chen, C. T. Au, Y. W. Du, Journal of Magnetism and Magnetic Materials, **261,** 238 (2003).
30. Nguyen Chau, Hoang Nam Nhat, Nguyen Hoang Luong, Dang Le Minh, Nguyen Duc Tho, Nguyen Ngoc Chau, Physica B **327**, 270 (2003).
31. Soma Das, T.K. Dey. Solid State Communications. **134**, 837 (2005).
32. S. K. Srivastava and S.Ravi, Journal of Physics: Condensed Matter **20**, 505212 (2008).
33. A.M. Tishin, K.A. Gschneidner, and V. K. Pecharsky, Phys. Rev. B **59**, 503 (1999).
34. Manjusha Battabyal, T.K. Dey, Physica B **367**, 40 (2005).
35. Wei Zhong, Wei Chen, Weiping Ding, Ning Zhang, Youwei Du, Qijie Yan, Solid State Communications **106**, 55 (1998).
36. C. Martin, A. Maignan, M. Hervieu, and B. Raveau, Phys. Rev. B **60**, 12191 (1999).
37. Lide M. Rodriguez-Martinez and J. Paul Attfield, Phys. Rev. B **54**, R15622 (1996).
38. H. Oesterreicher and F.T. Parker, Journal of Applied Physics, **55**, 4334 (1984).
39. V. Franco, J.S. Blarguez, and A. Conde, Applied Physics Letters, **89**, 222512 (2006).
40. A.G. Gamzatov, A.B. Batdalov, A.M. Aliev, L.N. Khanov, H. Ahmadvand, H. Salamati, P. Kameli, JETP Letters **91**, 341 (2010).

Table 1. Unit cell parameters of $La_{1-x}K_xMnO_{3+\delta}$ solid solutions at room temperature.

| $x$ in $La_{1-x}K_xMnO_{3+\delta}$ | hexagonal unit cell | | | rhombohedral unit cell | |
|---|---|---|---|---|---|
| | $a$, Å | $c$, Å | $V$, Å$^3$ | $a$, Å | $\alpha$, degrees |
| 0,000 | 5,5256(3) | 13,3384(9) | 352,69(3) | 5,4722(3) | 60,646(6) |
| 0,050 | 5,5204(3) | 13,3565(9) | 352,51(3) | 5,4754(3) | 60,544(6) |
| 0,100 | 5,5212(3) | 13,3750(9) | 353,10(3) | 5,4807(3) | 60,489(6) |
| 0,110 | 5,5191(3) | 13,3788(9) | 352,93(3) | 5,4810(3) | 60,460(6) |
| 0,130 | 5,5171(3) | 13,3857(9) | 352,86(3) | 5,4822(3) | 60,422(6) |
| 0,150 | 5,5102(3) | 13,3831(9) | 351,91(3) | 5,4792(3) | 60,375(6) |
| 0,175 | 5,5064(3) | 13,3849(9) | 351,47(3) | 5,4784(3) | 60,338(6) |

Table 2. Chemical analysis for samples of nominal composition $La_{1-x}K_xMnO_{3+\delta}$

| Nominal composition | | Determinated composition | |
|---|---|---|---|
| La/Mn | K/Mn | La/Mn | K/Mn |
| 1.000 | 0.000 | 0,995(8) | — |
| 0.950 | 0.050 | 0,947(8) | 0,048(8) |
| 0.900 | 0.100 | 0,896(6) | 0,095(6) |
| 0.890 | 0.110 | 0,891(7) | 0,112(8) |
| 0.870 | 0.130 | 0,870(8) | 0,124(9) |
| 0.850 | 0.150 | 0,850(6) | 0,146(9) |
| 0.825 | 0.175 | 0,825(6) | 0,173(6) |
| 0.800 | 0.200 | 0,803(6) | 0,192(9) |

Table 3. Some characteristics of the investigated samples.

| $x$ | $T_{max}$, K ($C_P$) | $T_{max}$, K (d$C_P$/d$T$) | $T_{max}$, K (MCE) | $\Delta T$ $\Delta H$=500 Oe | $\Delta S$, J/kg K $\Delta H$=500 Oe |
|---|---|---|---|---|---|
| 0.05 | - | - | 186.3 | 0.021 | 0.042 |
| 0.10 | 271.62 | 274.77 | 274.5 | 0.089 | 0.185 |
| 0.11 | 287.90 | 291.82 | 287.9 | 0.101 | 0.197 |
| 0.13 | 308.46 | 310.03 | 309.9 | 0.136 | 0.262 |
| 0.15 | 324.67 | 327.11 | 327.9 | 0.108 | 0.194 |
| 0.175 | 332.84 | 337.68 | 337.0 | 0.067 | 0.117 |

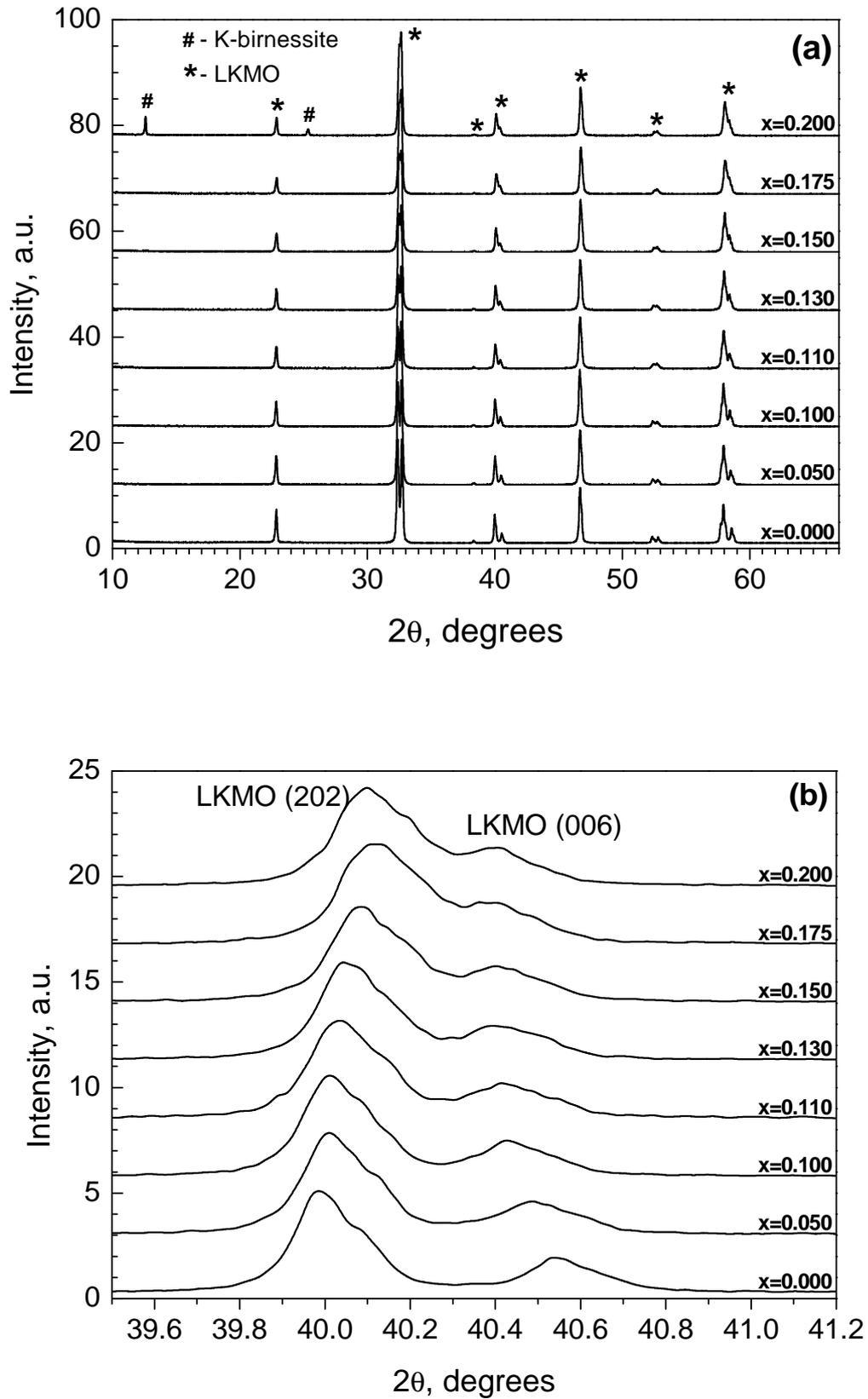

Fig. 1. Powder XRD pattern of the solid solutions La$_{1-x}$K$_x$MnO$_{3+\delta}$ (0≤$x$≤0.2): (a)-general view, (b)-enlarged region of (111)-pseudocubic reflection, showed indices corresponded to hexagonal unit cell.

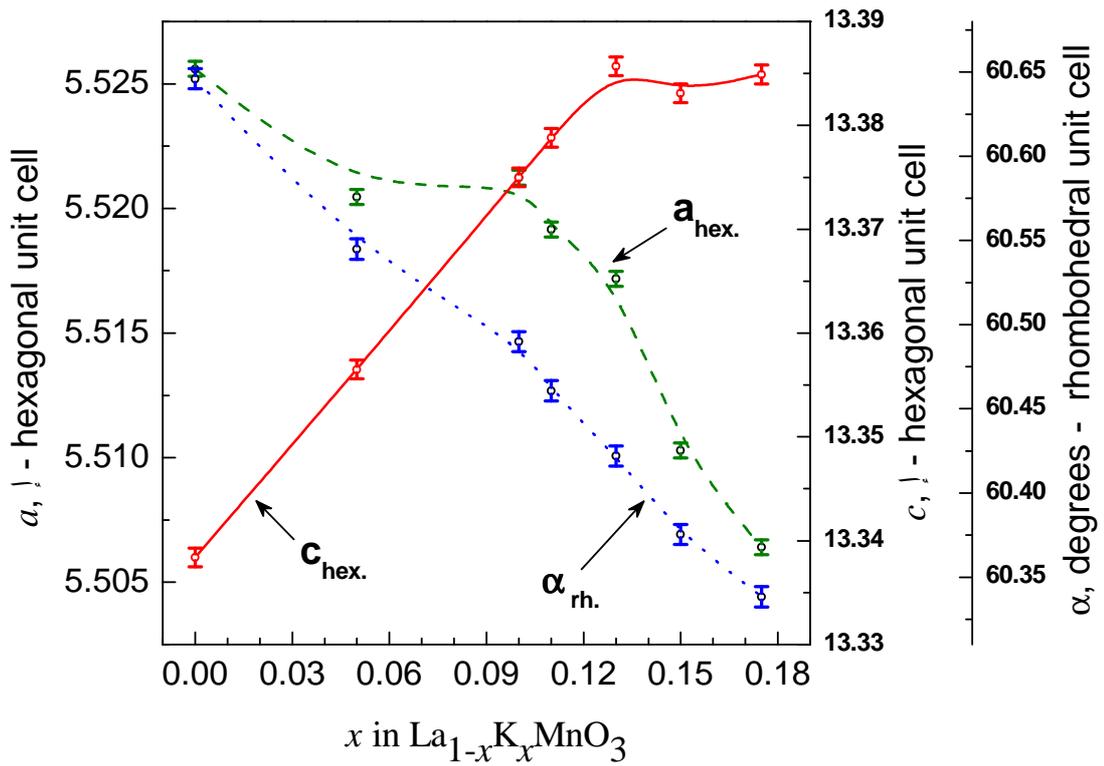

Fig. 2. Potassium content dependence of unit cell parameters of the $La_{1-x}K_xMnO_{3+\delta}$ solid solutions ($0 \leq x \leq 0.175$).

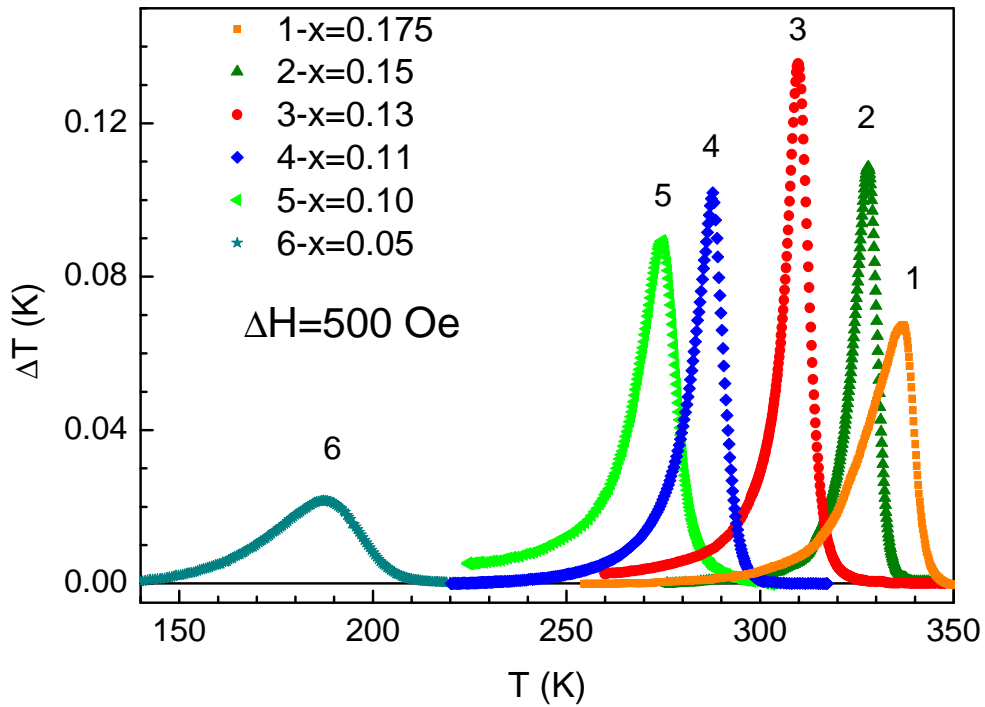

Fig. 3. Dependence of MCE for $La_{1-x}K_xMnO_3$ on the temperature with a change in the magnetic field $\Delta H$=500 Oe (modulation method)

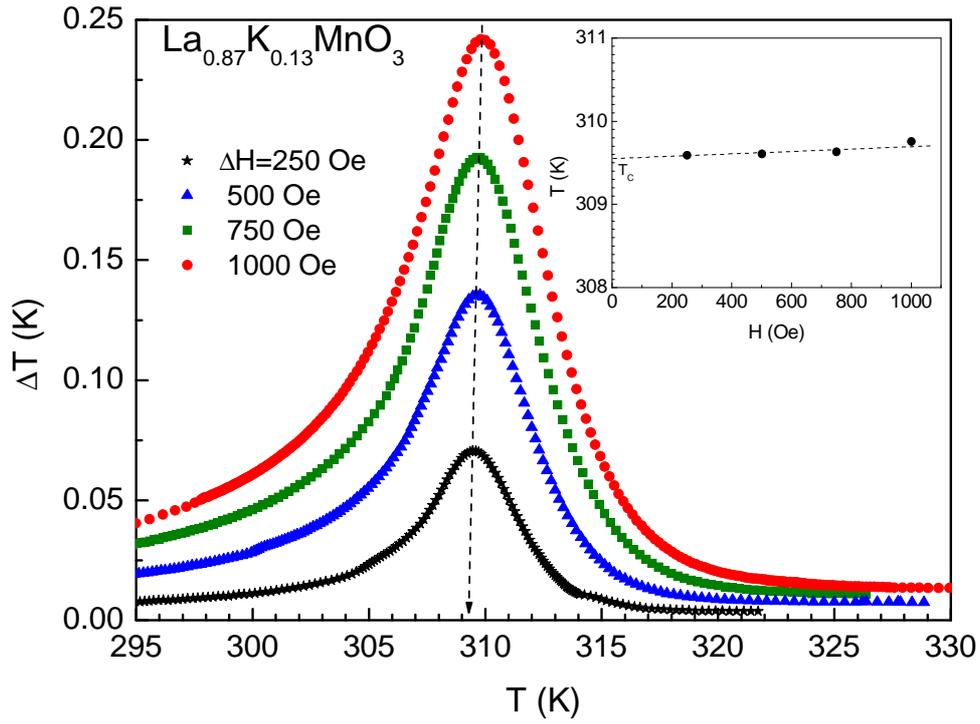

Fig. 4. Illustration of the $T_C$ determination by MCE data.

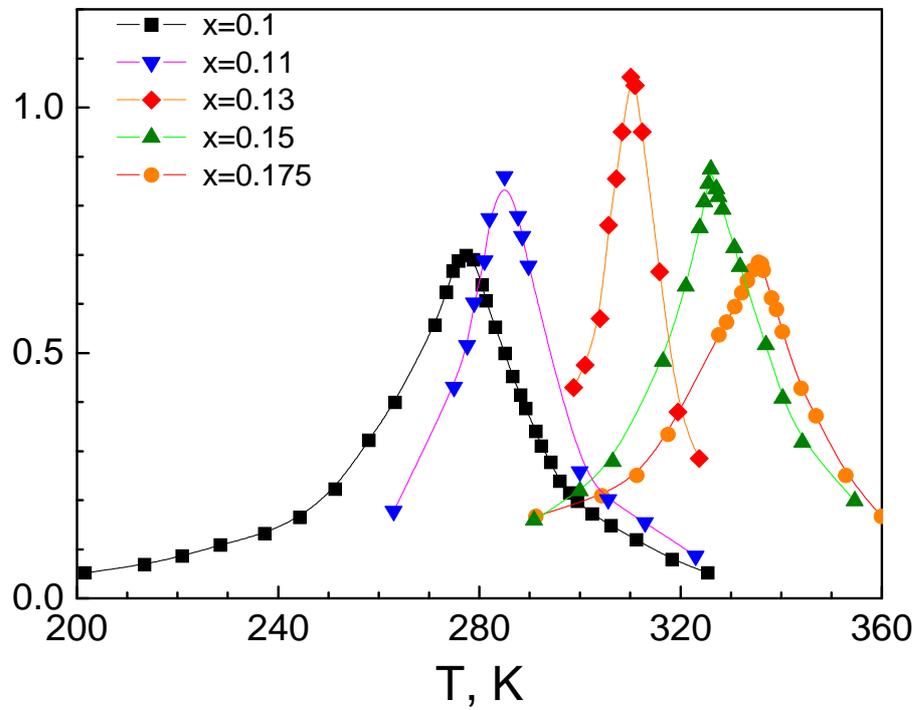

Fig. 5. MCE for the system of $La_{1-x}K_xMnO_3$ with $\Delta H$=11 k Oe (classical direct method).

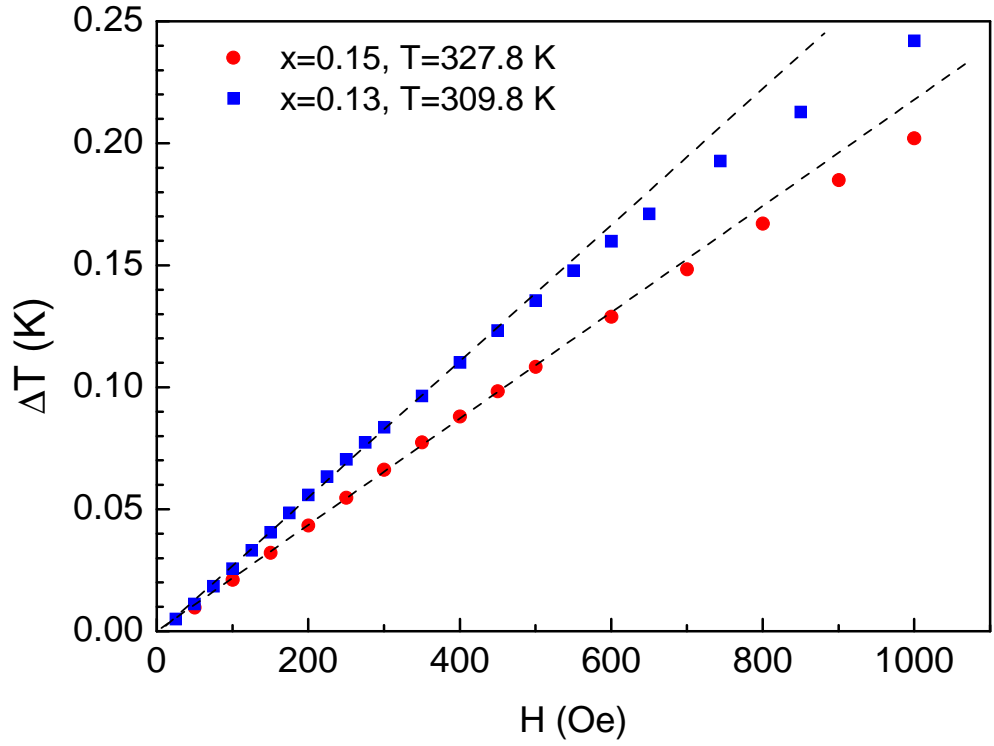

Fig. 6. Dependence of MCE for the samples $La_{0.87}K_{0.13}MnO_3$ and $La_{0.85}K_{0.15}MnO_3$ on the magnetic field near $T_C$. Dotted lines mean the linear extrapolation.